\documentclass[aip,floatfix,amsfonts,amssymb,amsmath,preprint]{revtex4-1}

\usepackage{graphicx}
\usepackage{dcolumn}
\usepackage{bm}

\begin{document}

\preprint{AIP/123-QED}

\title{Phonon Anharmonicity in Bulk $T_d$-MoTe$_2$}

\author{Jaydeep Joshi}
\author{Iris Stone}
\affiliation{Department of Physics and Astronomy, George Mason University, Fairfax, VA 22030, USA.}
\author{Ryan Beams}
\author{Sergiy Krylyuk}
\affiliation{Material Measurement Laboratory, National Institute of Standards and Technology, Gaithersburg, MD 20899, USA.}
\affiliation{Institute for Research in Electronics and Applied Physics, University of Maryland, College Park, MD, 20742, USA.}
\author{Irina Kalish}
\author{Albert Davydov}
\affiliation{Material Measurement Laboratory, National Institute of Standards and Technology, Gaithersburg, MD 20899, USA.}
\author{Patrick Vora}
\email{pvora@gmu.edu}
\affiliation{Department of Physics and Astronomy, George Mason University, Fairfax, VA 22030, USA.}

\date{\today}

\newcommand{\mote}{MoTe$_2$}
\newcommand{\mos}{MoS$_2$}
\newcommand{\mose}{MoSe$_2$}
\newcommand{\rese}{ReSe$_2$}
\newcommand{\snse}{SnSe$_2$}
\newcommand{\ws}{WS$_2$}
\newcommand{\wse}{WSe$_2$}
\newcommand{\wte}{WTe$_2$}
\newcommand{\mwte}{Mo$_\text{x}$W$_\text{1-x}$Te$_2$}
\newcommand{\td}{$T_d$}
\newcommand{\tp}{$1T^\prime$}
\newcommand{\gman}{Gr{\"u}neisen}
\newcommand{\icm}{cm$^{\text{-1}}$}
\newcommand{\degc}{$^{\circ}$C}
\newcommand{\degrees}{$^{\circ}$}

\def\introfig{
		\begin{figure}[t]
			\centering
			\includegraphics[trim = 0in 0.05in 0.25in .3in, clip=true,width=3.375in]{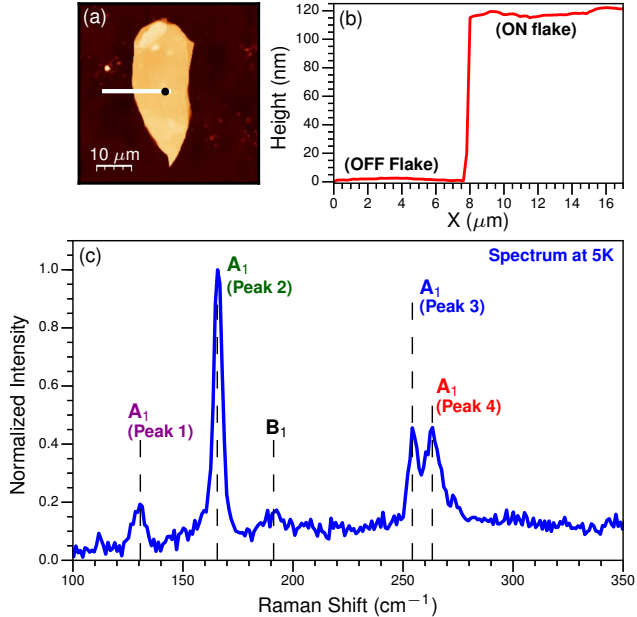}
			\caption{(a) AFM image of the \mote{} flake on a Si/SiO$_2$ substrate. The black circle on the flake marks the location where Raman spectra were acquired. (b) Height profile of the \mote{} flake extracted from (a) along the white line. The flake is over 100 nm thick, which corresponds to the bulk regime. (c) Raman spectra of \td{}-\mote{} acquired at 5 K. The four peaks examined in this study are labeled along with the symmetry assignments.}
			\label{introfig}
		\end{figure}
		}

\def\tempdepfig{
		\begin{figure}[t]
			\centering
			\includegraphics[trim = 0in 0.2in 0in .55in, clip=true,width=3.375in]{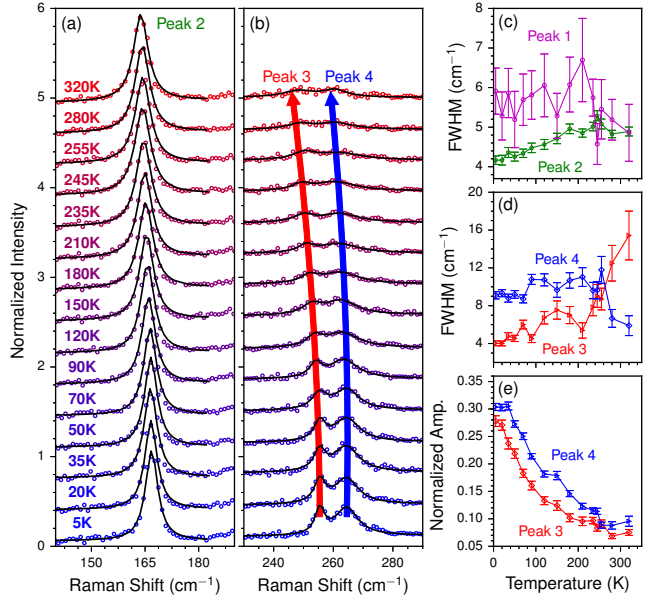}
			\caption{Temperature-dependent Raman spectra for (a) Peak 2 and (b) Peaks 3 and 4. Spectra are fit to either a single or double Lorentzian function with a linear background. The sample temperature increases from 5 K (bottom) to 320 K (top). Similar data for Peak 1 is presented in the Supporting Information. In both panels, the spectra have been normalized by the intensity of Peak 2. (c) FWHM of Peaks 1 and 2 extracted from the fits versus temperature. (d) FWHM of Peaks 3 and 4 extracted from the fits versus temperature. (e) Normalized amplitudes of Peaks 3 and 4 extracted from the fits in (b). $1\sigma$ error bars are included.}\label{tempdepfig}
		\end{figure}
		}

\def\fitfig{
		\begin{figure*}[t]
			\centering
			\includegraphics[trim = 0.3in 0.2in 0.6in 0.2in, clip=true,width=6.75in]{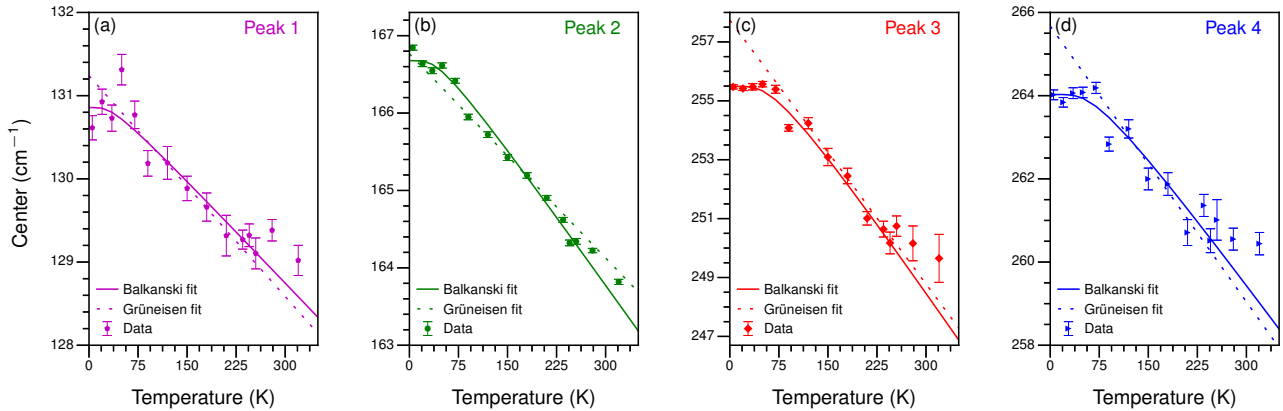}
			\caption{Raman frequency versus temperature for (a) Peak 1, (b) Peak 2, (c) Peak 3, and (d) Peak 4. In each panel, we fit the data to the \gman{} model (dashed lines) in Eq.~\ref{geq} from 100 - 200 K and to Eq.~\ref{baleq} (solid lines) from 0 - 250 K. The best fits are extrapolated over the entire temperature range and $1\sigma$ error bars are included.}\label{fitfig}
		\end{figure*}	
		}

\def\DataTable{
	\begingroup
	\squeezetable
		\begin{table*}[b]
			\caption{Measured values of $A$, $\omega_B$, $\chi$, and $\omega_o$ for Peaks 1-4 along with literature values for bulk $2H$-\mos{}.}
			\label{DataTable}
			\begin{ruledtabular}
					\begin{tabular}{c c c c c c c c}
						Mode & Material & Ref. & $A$ (\icm{}) & $\omega_B$ (\icm{}) & $\chi$ (\icm{}/K)\footnote{\label{fn1}Calculated from fits to data from 100 K to 200 K.} & $\omega_o$ (\icm{})$^\text{\ref{fn1}}$\\
						\hline\\
						Peak 1 & Bulk \td{}-\mote{} & this work & -0.390 $\pm$ 0.027 & 131.25 $\pm$  0.091 & -0.0089 $\pm$ 0.004 & 131.24 $\pm$ 0.537 \\
						Peak 2 & '' & '' & -0.717 $\pm$ 0.009  & 167.39 $\pm$ 0.022  & -0.0089 $\pm$ 0.001 & 166.77 $\pm$ 0.122 \\
						Peak 3 & '' & '' & -3.019 $\pm$ 0.094  & 258.51 $\pm$ 0.116  & -0.0312 $\pm$ 0.002 & 257.91 $\pm$ 0.303 \\
						Peak 4 & '' & '' & -2.052 $\pm$ 0.105  & 266.08 $\pm$ 0.140  & -0.0227 $\pm$ 0.003 & 265.74 $\pm$ 0.438 \\
						$A_{1g}$   & Bulk $2H$-\mos{} & \onlinecite{Su2014a} &-5.687 & - & -0.0197  & -\\
						$E_{2g}^1$ & " & " &-3.058 & - & -0.0221  & -
					\end{tabular}
				\end{ruledtabular}
		\end{table*}
	\endgroup
}

\begin{abstract}
We examine anharmonic contributions to the optical phonon modes in bulk \td-\mote{} through temperature-dependent Raman spectroscopy. At temperatures ranging from 100 K to 200 K, we find that all modes redshift linearly with temperature in agreement with the \gman{} model. However, below 100 K we observe nonlinear temperature-dependent frequency shifts in some modes. We demonstrate that this anharmonic behavior is consistent with the decay of an optical phonon into multiple acoustic phonons. Furthermore, the highest frequency Raman modes show large changes in intensity and linewidth near $T\approx 250$ K that correlate well with the \td$\to$\tp{} structural phase transition. These results suggest that phonon-phonon interactions can dominate anharmonic contributions at low temperatures in bulk \td-\mote{}, an experimental regime that is currently receiving attention in efforts to understand Weyl semimetals.\end{abstract}

\maketitle                             

\section{Introduction}
Transition metal dichalcogenides (TMDs) exhibit a wide range of layer-dependent phenomena depending on the choice of the transition metal and chalcogen atoms. Beginning with the discovery of photoluminescence from monolayer \mos{},\cite{Mak2010,Splendiani2010a} most studies of TMDs have focused on the semiconducting $2H$ ($\alpha$) crystal phase in \mos{}, \mose{}, \ws{}, \wse{}, and \mote{}.\cite{Yang2015,Gutierrez2013a,Chhowalla2013a,Wang2012c,Geim2013,Mak2016} The $2H$ structure is similar to the honeycomb lattice of graphene, but has broken sublattice symmetry and therefore a large bandgap (1-2 eV) as well as large exciton binding energies. However, unlike graphene and boron nitride, TMDs can exist in multiple structural phases that each exhibit unique electronic properties. \mote{} is one such TMD where the barrier between $2H$ and the alternative inversion-symmetric, monoclinic \tp{} ($\beta$) phase is minimal.\cite{Duerloo2014,Li2016} \tp{} \mote{} can be achieved by a modified growth method\cite{Keum2015, Park2015, Qi2015a,Naylor2016} as well as a low-temperature solution phase synthesis procedure\cite{Sun2016} and has recently also been found to be a low-bandgap semiconductor when in few-layer form.\cite{Keum2015} In addition, laser heating induces a structural transition from $2H$ to a metallic phase,\cite{Cho2015} which potentially enables \mote{} homojunctions as well as phase change memories.\cite{Duerloo2014}

A third crystal phase exists in \mote{} and \wte{} that is orthorhombic with broken inversion symmetry and is referred to as \td{}. For \wte{}, \td{} is thermodynamically stable at room temperature and pressure, whereas for \mote{} the \td{} phase only occurs at low temperature. The transition from \tp{}$\to$\td{} in \mote{} occurs near $T\approx 250$ K and can be observed in temperature-dependent electrical measurements as hysteresis after a warming/cooling cycle,\cite{Hughes1978,Zandt2007,Qi2015a} low-temperature Raman spectra through the activation of a inversion-symmetry forbidden shear mode,\cite{Chen2016} and X-ray diffraction (XRD) measurements.\cite{Clarke1978,Wang2016} The lattice constants in \td{}-\mote{} and \tp{}-\mote{} are only slightly different\cite{Dawson1987,Qi2015a,Wang2016} but nevertheless lead to striking modifications of the electronic structure, the most notable of which is the prediction of a type II Weyl semimetal phase.\cite{Sun2015a} This exciting electronic state has also been predicted in \wte{}\cite{Soluyanov2015} and the alloy \mwte{},\cite{Chang2015} and recent experimental results have confirmed the presence of the Weyl semimetal state in all three systems.\cite{Xu2016a,Belopolski2016,Wu2016,Wang2016b,Deng2016,Liang2016,Jiang2016,Huang2016,Tamai2016}

The clear importance of \td{}-\mote{} demands a systematic investigation of its properties and how they evolve under external stimuli. For example, high pressure has been shown to increase the superconducting transition temperature in \td{}-\mote{} substantially.\cite{Qi2015a} Thus far there has been little work on understanding how the vibrational properties evolve with temperature\cite{Chen2016} and the role of electron-phonon or phonon-phonon interactions in \tp{} or \td{}-\mote{}, both of which play important roles in the electronic properties of materials. Understanding the origin of anharmonic effects in \mote{} is therefore important for future explorations of low-temperature phenomena.

Here, we use temperature-dependent Raman spectroscopy to characterize anharmonic contributions to four prominent optical phonon modes in bulk \td{}-\mote{}. Our measurements extend down to 5 K, allowing us to observe departures from the commonly-observed \gman{} behavior. We find that two phonon modes exhibit modest changes in frequency with temperature, while the remaining two modes exhibit large, nonlinear changes in linewidth, frequency, and intensity. All four modes exhibit a change in frequency and slope around $T\approx 250$ K, which correlates well with the \td$\to$\tp{} structural phase transition. These observations are consistent with a regime where anharmonic contributions arising from optical phonon decay into multiple acoustic phonons are substantial.\cite{Balkanski1983}  Our results are the first studies of phonon anharmonicity in \td-\mote{} and provide crucial information for understanding the low-temperature electronic and vibrational properties of this highly-relevant material.
\introfig
\tempdepfig
\fitfig
\section{Experimental Methods}
\mote{} single crystals were produced by the chemical vapor transport (CVT) method with iodine as the transport agent. First, \mote{} powder was synthesized by annealing a stoichiometric mixture of molybdenum (99.999\%) and tellurium (99.9\%) powders at 750 \degc{} for 72 h in an evacuated and sealed quartz ampoule. To obtain \mote{} single crystals, approximately 2 g of polycrystalline \mote{} powder and a small amount of iodine (99.8\%, 4 mg/cm$^3$) were sealed in an evacuated quartz ampoule of 170 mm in length and 13 mm in diameter. The ampoule was placed in a horizontal furnace with a temperature gradient so that the end containing the \mote{} charge maintained a temperature of 1000 \degc{}, while the opposite end was kept at about 950 \degc{}. After 168 hours in the furnace, the ampoule was quenched in ice-water and the \mote{} single crystalline platelets were extracted. According to $\theta - 2\theta$ XRD scans, the platelets crystallized in the \tp{} form (space group $P2_1 /m$) with the lattice parameters $a=6.339(3)$ \AA, $b=3.466(4)$ \AA, $c=13.844(3)$ \AA, and $\beta=93.84(5)$\degrees{}.

Following structural characterization, the bulk \tp{}-\mote{} crystal was mechanically exfoliated using tape and deposited onto Si/SiO$_2$ substrates with an oxide thickness of 285 nm. Atomic force microscopy (AFM) images (Fig.~\ref{introfig}a) indicated that the flake we studied is $\approx 120$ nm thick (Fig.~\ref{introfig}b) and therefore lies well within the bulk regime, implying that interactions with the substrate can be neglected. It is notable that both the starting crystal and bulk flake we studied were exposed to atmosphere for prolonged periods of time, which is known to oxidize defects in $2H$-\mote{}.\cite{Chen2015a} A recent study demonstrated a similar effect in \tp{}-\mote{} where the Raman signal degraded during measurement, presumably due to photooxidation.\cite{Naylor2016} In order to avoid these effects, we performed all Raman experiments in vacuum and verified that our signal does not degrade after numerous cooldowns and measurements.

Raman measurements were performed from 5 K to 320 K on a home-built, confocal microscope integrated with a closed-cycle cryostat. The excitation source was a 532 nm laser focused through a 0.5 NA long working distance objective. Since the sample was excited in a backscattering geometry, we could detect Raman modes with $A_g$ and $B_g$ symmetry for \tp{} and $A_1$ and $B_1$ symmetry modes for \td{}. The laser spot diameter was $\approx 1.5$ $\mu$m at the sample and the laser power was 2 mW pre-objective for all measurements. Raman scattering from the sample was directed to a 500 mm focal length spectrometer with a LN$_2$ cooled CCD. The spectrometer and camera were calibrated using a Hg-Ar atomic line source and the spectrometer grating was positioned so that each Raman spectrum included the filtered laser line, allowing us to easily account for any drift in the laser wavelength. The instrumental response function (IRF) for these measurements was $\approx 3.9$ \icm{}.

\section{Results and Discussion}
We present an example Raman spectrum acquired at 5 K in Fig.~\ref{introfig}c, at which point \mote{} is in the \td{} phase. We observe four prominent peaks at $\approx$ 130 \icm{}, 165 \icm{}, 254 \icm{}, and 264 \icm{}, all with $A_1$ symmetry, which we refer to as Peaks 1-4, respectively. There is also a weak feature at 190 \icm{} that has $B_1$ symmetry, but its signal is not sufficient for temperature-dependent analysis. These assignments are based on polarization-dependent Raman measurements of the \tp{} phase at 300 K, which will be reported elsewhere, and density functional theory calculations of the phonon eigenspectrum.\cite{Sun2016} The Raman tensors for the $A_g$($B_g$) modes and the $A_1$($B_1$) modes are identical in structure for the backscattering configuration used here, which implies that they will evolve into an $A_1$($B_1$) symmetry at low temperature (Supporting Information).

As we increase the sample temperature from 5 K to 320 K, all peaks soften although the magnitude of the redshift is different for each. In Fig.~\ref{tempdepfig}a we present temperature-dependent Raman spectra for Peak 2 along with Lorentzian fits (black curves). Fig.~\ref{tempdepfig}b shows similar data for Peaks 3 and 4, and all spectra in Fig.~\ref{tempdepfig} have been normalized by the intensity of Peak 2. A similar analysis of Peak 1 is presented in the Supporting Information. 

We find that the full width at half maximum (FWHM) linewidth of Peak 1 does not change substantially over the entire temperature range (Fig.~\ref{tempdepfig}c). The FWHM of Peak 2 is limited by the IRF and also exhibits modest temperature dependence, which suggests that phonon-phonon interactions are weak for these modes. Peaks 3 and 4 exhibit markedly different behavior. As the sample temperature increases, these modes both redshift substantially and the FWHM linewidth of Peak 3 broadens (Fig.~\ref{tempdepfig}d). We quantify this behavior by fitting Peaks 3 and 4 to a double Lorentzian function plus a linear background, which allows us to extract the center frequency, amplitude, and linewidth of each mode.  At 5 K, the IRF limits the FWHM of Peak 3, which is narrower than Peak 4 ($\approx 4$ \icm{} versus $\approx 9$ \icm{}). However, Peak 3 steadily broadens with increasing temperature from 5 K to 250 K, at which point the two modes have similar linewidths. Over the same temperature range, the amplitude of Peak 4 decreases, becoming comparable to that of Peak 3 by 250 K (Fig.~\ref{tempdepfig}e). It is notable that observed changes in peak intensity and linewidth correlate well with the \td{}$\to$\tp{} transition temperature,\cite{Hughes1978,Zandt2007,Qi2015a,Chen2016,Wang2016} suggesting that the relative intensity and linewidth of Peaks 3 and 4 are sensitive to the change in structural phase.

The frequency of all examined Raman peaks softens with increasing temperature (Fig.~\ref{fitfig}). Temperature-induced shifts in Raman frequencies are typically fit using the linear \gman{} model:\cite{Jana2015,apinska2016,Pawbake2016,Late2014}

\begin{equation}
\omega(T) = \omega_o +\chi T.
\label{geq}
\end{equation}

\noindent{}$\omega (T)$ is the temperature-dependent phonon frequency, $\omega_o$ is the harmonic phonon frequency at 0 K, and $\chi$ is the first order temperature coefficient. The \gman{} model combines the effects of thermal expansion and the phonon self-energy into $\chi$ and is sufficient when $T >> \frac{\hbar\omega_o}{2k_B}$ where $k_B$ is the Boltzmann constant. At lower temperatures, $\omega (T)$ can become nonlinear and Eq.~\ref{geq} is not necessarily valid. We illustrate this explicitly in Fig.~\ref{fitfig}, where the dashed line is a fit of Eq.~\ref{geq} to the data from 100 K to 200 K. While there is good agreement at low $T<100$ K for Peaks 1 and 2 (Figs.~\ref{fitfig}a and \ref{fitfig}b), large deviations exist in this range for Peaks 3 and 4 (Figs.~\ref{fitfig}c and \ref{fitfig}d). The fact that the linewidths of Peaks 1 and 2 exhibit comparably little variation suggests that thermal expansion and/or electron-phonon coupling, rather than phonon-phonon interactions, dominate anharmonic effects in these modes. For Peaks 3 and 4, we suggest that thermal expansion of the lattice makes a negligible contribution to the phonon frequency, and instead changes in the phonon self-energy from anharmonic coupling between different phonon branches are the dominant effect.\cite{Su2014a} Theoretical work by Balkanski et al. demonstrated that the temperature dependence of optical phonons can be accounted for by a decay pathway resulting in multiple acoustic phonons.\cite{Balkanski1983} $\omega (T)$ in this formalism is given by

\begin{equation}
\omega(T) = \omega_B +A\Big(1+\frac{2}{e^x-1}\Big)
\label{baleq}
\end{equation}

\noindent{}where $x=\frac{\hbar\omega_B}{2 k_B T}$ and $\omega_B$ is the 0 K harmonic phonon frequency. In the treatment by Ref. \onlinecite{Balkanski1983}, the constant $A$ represents an anharmonic contribution to the frequency involving the decay of an optical phonon into two acoustic phonons. At $T=0$, $\omega(0) = \omega_B +A$, which implies that $A$ represents a third order correction to the phonon self-energy.\cite{Balkanski1983} This model has been successfully applied to a variety of nanomaterials to explain nonlinear temperature-dependence in optical phonon frequencies.\cite{Taube2015,Duzynska2014,apinska2016,Taube2014,Su2014a}

We fit the frequencies of Peaks 1-4 (Figs. \ref{fitfig}a - \ref{fitfig}d) to Eq.~\ref{baleq} up to $250$ K in order to isolate the \td{} crystal phase. The extracted values of $A$, $\omega_B$, $\chi$, and $\omega_o$ are summarized in Table \ref{DataTable} and compared to bulk $2H$-\mos{}.\cite{Su2014a} We find that Peaks 3 and 4 have larger values of $A$ which, when combined with the significant changes in FWHM, suggest that phonon interactions dominate the anharmonic coupling. Note that an additional term in Eq.~\ref{baleq} corresponding to fourth order processes has been omitted as it was not required to fit our data.\cite{Balkanski1983} Furthermore, we observe that all four modes deviate from both Eqs.~\ref{geq} and \ref{baleq} above $T\approx 250$ K, which we interpret as the onset of the \tp{} phase with distinct anharmonic behaviors.
\DataTable
The unique behavior of Peaks 3 and 4 is attributed to the apparent sensitivity of their atomic displacements to interlayer coupling. As \td{} transitions to \tp{}, the $a$ and $b$ lattice parameters increase by 0.0198 \AA{} and 0.0375 \AA{}, respectively, while the $c$ lattice parameter decreases by 0.0447 \AA{}.\cite{Tamai2016} This implies that the dominant changes are in the directions of the $b$ lattice vector and the out-of-plane $c$ lattice vector. Furthermore, the angle between the $b$ and $c$ lattice vectors changes from 90\degrees{} in \td{} to 93.84\degrees{} in \tp{}.\cite{Tamai2016} These combined structural changes imply that phonon modes sensitive to interlayer coupling will be most affected by the \td{}$\to$\tp{} transition. Peak 2 depends only weakly on the number of layers,\cite{Keum2015} and therefore it is expected to respond minimally to changes in layer alignment and separation. In contrast, Peaks 3 and 4 are sensitive to flake thickness\cite{Keum2015} and are therefore likely to respond to changes in the spacing and alignment of the layers during the \td{}$\to$\tp{} transition.

\section{Conclusion}
In summary, we have performed the first study of optical phonon anharmonicity in bulk \td{}-\mote{}. Changes in phonon frequency, linewidth, and amplitude are determined for four modes and the results correlate well with the \td{}$\to$\tp{} structural phase transition. $\omega(T)$ is nonlinear for two modes, which is consistent with an anharmonic contribution arising from optical phonon decay into multiple acoustic phonons. The large changes in frequency and linewidth for these two modes as \td{}$\to$\tp{} indicates they are highly sensitive to interlayer separation and alignment, an observation that is consistent with prior characterizations of their dependence on layer number. All four modes exhibit a change in slope and an increase in frequency at $T\approx 250$ K which we attribute to the \td{}$\to$\tp{} structural phase transition. These results highlight the important, and in some cases dominant, role of phonon-phonon interactions in \td{}-\mote{}. Further studies connecting the atomic displacement of these modes to their anharmonicity are desirable and additional investigation of suspended, few-layer \mote{} will illuminate the impact of interlayer coupling on anharmonicity in this exciting material.

\section*{Supporting Information}
See supporting information for the analysis of Peak 1 and a discussion of the Raman tensors of \tp{} and \td{} \mote{}.

\begin{acknowledgments}
J.J., I.S., and P.M.V. acknowledge support from the Office of Naval Research through grant N-00014-15-1-2357 and from the GMU OSCAR Program. R.B. thanks the National Research Council Research Associateship Programs for its support.
\end{acknowledgments}

\bibliography{mote2_tempdep_apl_final_forarxiv}

\end{document}